\documentclass[aps,prb,twocolumn,superscriptaddress,floatfix,citeautoscript,preprintnumbers,noeprint]{revtex4-2}
\usepackage{graphicx,color}
\usepackage{amssymb}
\usepackage{amsmath}
\usepackage[normalem]{ulem}
\usepackage{comment}
\usepackage{breakcites}
\usepackage[breaklinks=true]{hyperref}
\usepackage[caption=false]{subfig}
\usepackage{enumitem}
\usepackage{soul}
\usepackage{bbold,dsfont}
\usepackage{braket}
\usepackage{makecell}
\usepackage{textcomp}
\usepackage{multirow}
\usepackage{titlesec}

\def\Id{\mathbb{1}}

\def\Z2{Z_{\mathrm{st2}}}

\newcommand{\dInt}{{\mathrm{d}}}

\newcommand{\varO}{{\mathcal{O}}}
\newcommand{\floor}[1]{\lfloor #1 \rfloor}

\newcommand{\tr}{\mathop{\mathrm{tr}}}

\newcommand{\nn}{{\nonumber}}

\newcommand{\beq}{\begin{equation}}
\newcommand{\eeq}{\end{equation}}
\newcommand{\bal}{\begin{align}}
\newcommand{\eal}{\end{align}}

\newcommand{\beqa}{\begin{eqnarray}}
\newcommand{\eeqa}{\end{eqnarray}}

\newcommand{\Pd}{\hat{P}_{\delta}}  %P_delta
\newcommand{\Fd}{\hat{F}_{\delta,\alpha}}  %P_delta
\newcommand{\rhoF}{\rho_{(E,\delta)}}  %P_delta
\newcommand{\rhoFe}[1]{\rho_{(#1,\delta)}}

\definecolor{MyDarkBlue}{rgb}{0,0.08,0.45}
\definecolor{MyLightMagenta}{cmyk}{0.1,0.8,0,0.1}
\definecolor{MLM}{cmyk}{0.1,0.8,0,0.1}
\definecolor{MyDarkGreen}{rgb}{0,0.45,0.08}
\definecolor{MDG}{rgb}{0,0.55,0.05}

\usepackage[textsize=footnotesize]{todonotes}

\definecolor{atomictangerine}{rgb}{1.0, 0.6, 0.4}
\definecolor{bluegray}{rgb}{0.4, 0.6, 0.8}
\definecolor{brightube}{rgb}{0.82, 0.62, 0.91}
\definecolor{brilliantlavender}{rgb}{0.96, 0.73, 1.0}

\begin{document}

%\title{Finite temperature simulation through spectral filtering}
\title{Classical algorithms for many-body quantum systems at finite energies}

% ====================================================================================
%           ABSTRACT
% ====================================================================================

\begin{abstract}
    %	Energy filters provide a way to explore spectral properties of quantum many-body problems and allow access to microcanonical and diagonal ensembles. The hybrid algorithms recently proposed in [Lu et al. PRX Quantum 2, 020321 (2021)] demonstrate that with a quantum simulator it is possible to compute these quantities efficiently and to use them to investigate microcanonical and thermal equilibrium properties. Here we present a classical simulation algorithm that, inspired on the quantum protocol, allows us to compute such properties for much larger systems than exact diagonalization or previously proposed algorithms. This is especially interesting for microcanonical expectation values, for which very few other methods exist, in the general case. 	We discuss different options for the implementation of our method with tensor networks, and demonstrate its performance with numerical results obtain for spin chains up to 80 sites.
    We investigate quantum inspired algorithms to compute physical observables of quantum many-body systems at finite energies. They are based on the quantum algorithms proposed in [Lu et al. PRX Quantum 2, 020321 (2021)], which use the quantum simulation of the dynamics of such systems, as well as classical filtering and sampling techniques. Here, we replace the quantum simulation by standard classical methods based on matrix product states and operators. As a result, we can address significantly larger systems than those reachable by exact diagonalization or by other algorithms. We demonstrate the performance with spin chains up to 80 sites.
\end{abstract}

\author{Yilun Yang}
\affiliation{Max-Planck-Institut f\"ur Quantenoptik, Hans-Kopfermann-Str.\ 1, D-85748 Garching, Germany}
\affiliation{Munich Center for Quantum Science and Technology (MCQST), Schellingstr. 4, D-80799 M\"unchen}
\author{J. Ignacio Cirac}
\affiliation{Max-Planck-Institut f\"ur Quantenoptik, Hans-Kopfermann-Str.\ 1, D-85748 Garching, Germany}
\affiliation{Munich Center for Quantum Science and Technology (MCQST), Schellingstr. 4, D-80799 M\"unchen}
\author{Mari Carmen Ba\~nuls}
\affiliation{Max-Planck-Institut f\"ur Quantenoptik, Hans-Kopfermann-Str.\ 1, D-85748 Garching, Germany}
\affiliation{Munich Center for Quantum Science and Technology (MCQST), Schellingstr. 4, D-80799 M\"unchen}

\date{\today}							
\maketitle

% ====================================================================================
%          INTRODUCTION
% ====================================================================================

\section{Introduction}
\label{sec:intro}

Computing the properties of quantum systems in equilibrium is of central interest in many-body physics. For a system at finite temperature, there exists a wide spectrum of techniques that are used in practice. For large systems, where exact solutions are unreachable, they typically approximate the corresponding Gibbs state using variational, sampling, or series expansion methods~\cite{Sandvik1991, Verstraete2004, Rigol2006, Hastings2007a, White2009, Stoudenmire2010, Molnar2015}. For systems at a finite energy, e.g. in the microcanonical ensemble, methods are more scarce~\cite{Schrodi2017, Yang2020, Lu2021}.

A possible approach consists in simulating the dynamics and extracting the equilibrium properties from there. For instance, one can use spectral filters in order to retrieve expectation values of an observable $O$, by averaging them at different times. In this way, one obtains results connected to the diagonal ensemble corresponding to the initial state~\cite{Rigol2008a}, namely
\beq
\label{Oexp}
\bar O = \sum_n |c_n|^2 \langle E_n|O|E_n\rangle
\eeq
where $|E_n\rangle$ are the energy eigenstates and $c_n$ the coefficients of the initial state in that basis. For local Hamiltonians and initial states with finite correlation length (like product states), the values of $c_n$ are significant if $|E_n-E|<\delta =\varO(\sqrt{N})$, where $E$ is the mean energy of the initial state. Under the eigenstate thermalization hypothesis (ETH)~\cite{Deutsch1991,Srednicki1994,DAlessio2016}, $\bar O$ converges to the equilibrium value in the thermodynamic limit, and it is not necessary to average the results at different time but just to wait for a sufficiently long time. Quantum computers and analog quantum simulators are very well suited for that task, since they can deal with the dynamics of many-body quantum systems in a very natural way~\cite{Lloyd1996, Haah2018}. Classical methods to simulate the dynamics typically suffer from the linear growth of entanglement~\cite{Calabrese2005}, which gives an exponential cost with time. Even if the (weak) ETH applies, the thermalization time can be very long~\cite{Lan2018a, Lin2020}, and this severely restricts the applicability of such classical algorithms.

In this work we propose and analyze a classical algorithm to compute expectation values of the form~\eqref{Oexp}. In particular, for $c_n$ that are Gaussian functions of $|E_n-E|$ with a variance $\delta$, this can be achieved by simulating the dynamics for a time $\varO(1/\delta)$. This allows us, for instance, to reach $\delta=\varO(\sqrt{N})$ by just using standard time-evolution techniques for tensor networks for very short times, when the entanglement is still very small and thus the techniques work well. One can also reach values of $\delta=\varO(1)$ with modest computational resources. The algorithm is inspired by a quantum algorithm presented in~\cite{Lu2021} that allows one to compute expectation values of the form~\eqref{Oexp}. This algorithm combines classical sampling (Monte Carlo) techniques with time series and Loschmidt echo-like measurements~\cite{Peres1984, Jalabert2001} that can be obtained by quantum simulation of the dynamics. Our main modification is to replace the latter by a classical simulation using tensor network states~\cite{Verstraete2008, Schollwock2011, Huckle2013, Orus2014, Silvi2019, Okunishi2021}. This allows us to compute~\eqref{Oexp} for times $\varO(1/\delta)$ instead of the thermalization time, thus circumventing the problem of entanglement growth. This is done at the expense of having to sample, which just involves the repetition of the whole procedure until convergence. We apply the algorithm to one dimensional systems, sample over a basis of product states and use matrix product states (MPS) and operators (MPO) to simulate the evolution~\cite{Paeckel2019tevol}. We illustrate the performance of the method for a non-integrable Ising chain, for which we obtain convergence to the microcanonical values for systems up to $80$ sites, far larger than what is possible with exact diagonalization.

Apart from that, in ~\cite{Lu2021} another quantum algorithm was proposed to compute physical observables in a state where an energy filter of width $\delta$ is applied. For local Hamiltonians, the computational time also scales as $\varO(1/\delta)$. Here we also analyze a classical algorithm inspired in that method. We notice that this method typically requires  much narrower $\delta$ (thus longer times) to approach thermodynamic quantities. However, the filtering achieved with a limited evolution time can be optimized if the initial state is chosen with already a reduced energy width. Here we demonstrate this possibility by applying the classical version of the first algorithm on matrix product states found by minimizing the energy variance.

The rest of the paper is organized as follows. In section~\ref{sec:filters} we briefly review the concept of energy filters, introduce the filter ensemble and discuss its applications to determine microcanonical and diagonal properties. We also review briefly the quantum algorithms~\cite{Lu2021} that motivate this work. In section~\ref{sec:classical} we discuss the details of a TNS simulation of the quantum algorithms, the different possibilities and associated parameter choices. Section~\ref{sec:results} presents our numerical results for Ising chains, for each of the algorithms implemented. The paper is closed with the discussion in section~\ref{sec:discussion}.

% ====================================================================================
%          FILTERS
% ====================================================================================

\section{Filters and quantum algorithms}

We start by recalling the definition and properties of the energy filters that are at the basis of the algorithms in~\cite{Lu2021} and this work.

\label{sec:filters}
\subsection{Energy filters}
The main tool used by the finite energy algorithms discussed here is a filtering operator that suppresses energy eigenstates outside a target energy interval. In particular, given the Hamiltonian $H$, we define a Gaussian filter centered in energy $E$ and of width $\delta$ as the following operator
\begin{eqnarray}
    \Pd (E) = \exp \left[ - \left(\hat{H} - E \right)^2 / 2 \delta^2 \right].
    \label{eq:filterGauss}
\end{eqnarray}

Notice that, up to normalization, $\Pd (E)$ is a diagonal ensemble in the energy basis. We refer to it as the \emph{filter ensemble}
\beq
\rhoF=\frac{\Pd(E)}{\mathrm{tr}\left[\Pd(E)\right]}.
\label{eq:rhoF}
\eeq
The corresponding expectation values are precisely of the form~\eqref{Oexp}, with coefficients $|c_n|^2$ distributed according to a Gaussian of width $\delta$. For local Hamiltonians and large systems, product states have that kind of spectral decomposition, with $\delta\sim \varO(\sqrt{N})$~\cite{Hartmann2005,Keating2015} but, since they contain coherences in the energy basis, the corresponding expectation values are very different. An ensemble as~\eqref{eq:rhoF} could nevertheless be obtained from a product state, but only after evolving and averaging over a long time.

As the width $\delta$ is reduced, the filter approaches the microcanonical ensemble. Thus we can make use of the filter to access the microcanonical properties of the quantum system in the following two different ways.

\paragraph{Filtering a state.}
Given a state $\ket{\psi}$, its local density of states (LDOS) is defined as $D_{\psi}(E) = \braket{\psi | \delta(E - \hat{H}) | \psi}$. A broadened version can be computed with the filter as
\begin{eqnarray}
    D_{\delta,\psi}(E) = \frac{1}{\sqrt{2\pi} \delta} \braket{\psi | \hat{P}_{\delta} (E) |\psi}.
    \label{eq:LDOS}
\end{eqnarray}

We can also use the filtered state to explore the microcanonical ensemble expectation value $O_{\mathrm{micro}}(E)$ of an observable $\hat{O}$. In generic cases in which ETH is satisfied, and for a value $E$ at which the LDOS does not vanish,
\begin{eqnarray}
    O_{\delta}\left(E, \ket{\psi}\right) = \frac{\braket{\psi | \Pd (E) \hat{O} \Pd (E) | \psi }}{ \braket{\psi | \Pd (E)^2 | \psi} }
    \label{eq:micro_filter_state}
\end{eqnarray}
will converge to $O_{\mathrm{micro}}(E)$ in the limit $\delta \to 0$.

\paragraph{Filtering the whole spectrum.}
The filter ensemble itself converges to the microcanonical ensemble as the width is reduced. Hence, we can also use it without specifying a state, but directly taking its trace. In particular, the density of states (DOS) of the Hamiltonian $H$, defined as $D(E) = \tr \left[ \delta(E - \hat{H}) \right]$, can be approximated through the broadening of the $\delta$ functions as
\begin{eqnarray}
    D_{\delta}(E) = \tr \left[ \Pd (E) \right].
\end{eqnarray}
% The canonical ensemble partition function can then be obtained with a Laplace transform $Z_{\delta} (\beta)= \int \dInt E e^{-\beta E} D_{\delta}(E)$ and thus thermodynamic quantities can follow.

Moreover, the expectation values in the filter ensemble
\begin{eqnarray}
    O_{\delta} (E) =  \tr\left[ \hat{O}  \rhoF \right]=\tr \left[ \hat{O} \Pd (E) \right] / \tr \left[ \Pd (E)\right]
    \label{eq:micro_filter_tr}
\end{eqnarray}
will converge to the microcanonical values as $\delta \to 0$.

Whereas in generic cases one can in principle approach the microcanonical expectation values with either~\eqref{eq:micro_filter_state} or ~\eqref{eq:micro_filter_tr}, the convergence of the second with $\delta$ is much faster, since the filter ensemble is diagonal in the energy basis, while ~\eqref{eq:micro_filter_state} contains contributions from off-diagonal matrix elements~\cite{Beugeling2015, Luitz2016, Mondaini2017}, which can converge much slower than the diagonal part~\cite{DAlessio2016,Dymarsky2019}.

\begin{figure}[t]
    \includegraphics[width=.48\textwidth]{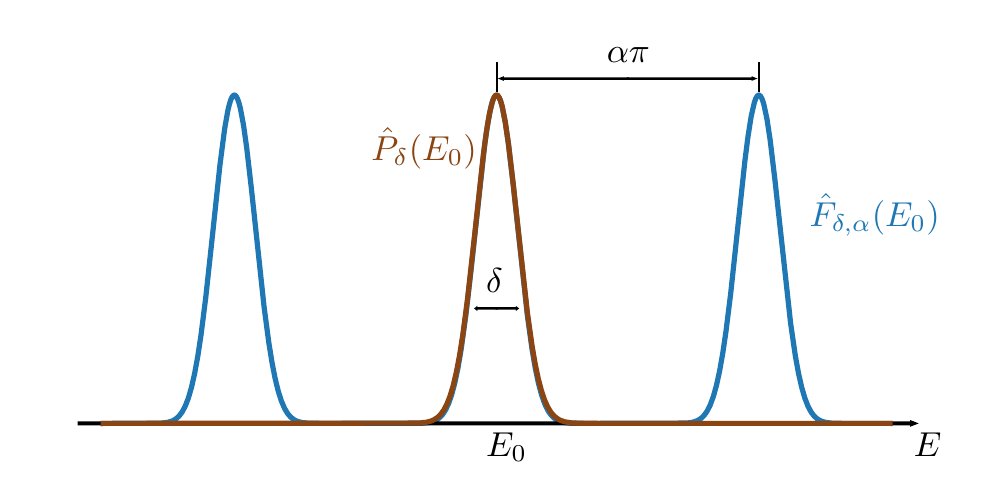}
    \caption{Approximating $\Pd(E_0)$ with $\Fd(E_0)$.}
    \label{fig:period}
\end{figure}

\subsubsection*{Implementing the filter}

For the purpose of numerical, but also quantum, simulations of the filter, it is convenient to substitute an approximation for the Gaussian filter. In this paper, following the quantum algorithms in~\cite{Lu2021}, we focus on the cosine filter~\cite{Banuls2019,Ge2019} defined as
\begin{eqnarray}
    \Fd (E) = \cos \left[ \left(\hat{H} - E\right) / \alpha \right]^{M} \approx \Pd (E),
    \label{eq:filter_cos}
\end{eqnarray}
where $\alpha$ is a parameter (with dimensions of energy) that controls the validity of the approximation, and $M = \floor{\alpha^2 / \delta^2}_2$ with $\floor{\cdots}_2$ giving the closest smaller even integer. The approximation is valid if $\left\lVert  \hat{H} - E \right\rVert  \le \alpha \pi / 2$, when the spectrum of $\hat{H}$ lies in one period of the cosine function~\footnote{The approximation actually works beyond the range, see Appendix E of \href{https://link.aps.org/doi/10.1103/PRXQuantum.2.020321}{PRX Quantum \textbf{2}, 020321 (2021)}.}. \eqref{eq:filter_cos} can be further approximated by a truncated series of evolution operators
\begin{eqnarray}
    \Fd (E) \approx \Fd^x (E) = \sum_{m = - R}^{R} c_m e^{-i\left(\hat{H} - E\right) t_m},
    \label{eq:filter_cos_sum}
\end{eqnarray}
where $R = \floor{x \alpha / \delta}$, $t_m = 2m / \alpha$, $x$ is a constant that bounds the truncation error in operator norm as $\left\lVert \Pd(E) - \Fd(E) \right\rVert \le 2e^{-x^2/2}$ and
\begin{eqnarray}
    c_m = \frac{1}{2^M} \begin{pmatrix}
        M \\ M /2 - m
    \end{pmatrix}.
\end{eqnarray}
With the cosine filter the problem is turned into evolving states or computing the traces of time evolution operators, which leads itself to a natural implementation in quantum simulators.

\subsection{The quantum algorithms}
\label{subsec:q_alg}
In Lu et al.'s paper~\cite{Lu2021}, two hybrid classical-quantum algorithms corresponding to the two different ways of applying the filter were introduced. We sketch them here for completeness.

The first one computes \eqref{eq:micro_filter_state} for a state $\ket{\psi}$ that can be easily prepared. Suppose the quantum device can efficiently obtain the following quantities
\begin{eqnarray}
    \begin{aligned}
        a_{\psi}(t)           & = \braket{\psi | e^{-i\hat{H}t} | \psi}                          \\
        a_{O, \psi}(t_1, t_2) & = \braket{\psi |e^{i\hat{H}t_1} \hat{O}  e^{-i\hat{H}t_2}|\psi},
    \end{aligned}
    \label{eq:a_psi}
\end{eqnarray}
then~\eqref{eq:micro_filter_state} can be determined by classical postprocessing as
\begin{eqnarray}
    O_{\delta}(E, \ket{\psi}) = \frac{\sum_{m,n=-R}^R  c_m^* c_n a_{O, \psi} (t_m, t_n) }{\sum_{m,n=-R}^R  c_m^* c_n a_{\psi} (t_n - t_m)},
\end{eqnarray}
without explicitly preparing the filtered state. The required time scale is proved to be a polynomial of system size $N$, the inverse of the width of filter $1 / \delta$ and the inverse of the error, provided the state $\ket{\psi}$ can be prepared efficiently, for a value of $E$ in a small interval around the mean energy of $\ket{\psi}$.

In the second algorithm (quantum-assisted Monte Carlo), importance sampling is applied to compute~\eqref{eq:micro_filter_tr}. Let us rewrite that expression as
\begin{eqnarray}
    O_{\delta}(E) = \frac{ \int \dInt \mu_{\phi} D_{\delta,\phi}(E) O_{\delta,\phi} (E) }{\int \dInt \mu_{\phi} D_{\delta,\phi}(E) },
\end{eqnarray}
where $\set{\ket{\phi}}$ is an (over-)complete basis, with $\dInt \mu_{\phi}$ the appropriate measure to ensure the closure relation $\int \dInt \mu_{\phi} \ket{\phi} \bra{\phi} = \Id$ (a simple choice is for instance the computational basis).
%where $\dInt \mu_{\phi}$ is a measure an (over-) complete basis set $\set{\ket{\phi}}$ fulfilling $\int \dInt \mu_{\phi} \ket{\phi} \bra{\phi} = \Id$, for instance the computational basis, 
$D_{\delta,\phi}$ is the LDOS defined in~\eqref{eq:LDOS} and
\begin{eqnarray}
    \quad O_{\delta,\phi} = {\braket{\phi | \hat{O} \Pd(E) | \phi}} / {\braket{\phi |  \Pd(E) | \phi}}.
    \label{eq:O_filter_state}
\end{eqnarray}
Both $D_{\delta,\phi} (E)$ and $O_{\delta,\phi}(E)$ can be obtained by measuring the quantities defined in~\eqref{eq:a_psi}, as long as we can run the first algorithm with the quantum device for the states in the basis $\set{\ket{\phi}}$. Then a Metropolis-Hastings step can be applied classically with regard to the probability distribution $D_{\delta,\phi}(E) / \int \dInt \mu_{\phi} D_{\delta,\phi}(E)$, and the value of $O_{\delta}(E)$ can be estimated. Because $D_{\delta,\phi}(E)$ is positive, this method does not encounter a sign problem.

Given $\delta$, this second algorithm provides access to observables in the filter ensemble at the cost of simulating time evolutions for times $\varO(1/\delta)$, at the expense of repeating the procedure until the sampling converges. This is especially remarkable because one could obtain a similar result from the time evolution of an initial state with the same distribution of coefficients, but this would require evolving for as long as the thermalization time. As a particular application, if one chooses $\delta =o(\sqrt{N})$, the expectation values of intensive quantities will (under the ETH) already be equivalent to those in the Gibbs ensemble at the same mean energy (see \ref{subsec:para_width}), and therefore this algorithm is an inexpensive way of accessing thermal properties.

If we are interested in microcanonical expectation values, we can use either algorithm, but we need to reduce the width of the filter. Since the trace quantities converge faster in $\delta$ to the microcanonical values, a shorter evolution time is required with this second algorithm. In exchange, the procedure needs to be repeated over many states, to perform the classical Monte Carlo sampling.

\subsubsection*{Extreme values of energy}

Both algorithms above rely on the evolution of easily preparable states (a requirement for the single initial state in the first algorithm or the whole sampling basis in the second). A most practical choice is that of product states. The
mean energies of such states are contained in an extensive but generally restricted interval within the spectrum~\cite{Lieb1973}, such that values close to the edges of the spectrum may be out of reach.
As indicated in~\cite{Lu2021}, the accessible range of energies can be extended by considering larger sets of states. In particular, MPS can be used to circumvent the limitation.

For the first algorithm, the initial state  $\ket{\psi}$ can be found as an MPS with a small bond dimension such that its energy expectation value is close enough to $E$, as MPS serves as a good representation of the ground state and low-lying excited states. This can be done, for instance, by first finding the MPS  with that bond dimension and minimal energy ($E_{\min}$), and then changing its parameters until the desired energy $E > E_{\min}$ is reached. Another possibility is to find the MPS minimizing $(H-E)^2$~\cite{Lim2016exc,Yu2017excited}.
% Once the MPS is known, it is also possible to prepare it efficiently in a quantum simulator~\cite{Schon2005, Schon2007, Wei2022}.

For the second algorithm, a different basis for Monte Carlo sampling can be chosen, where we start from any state $\ket{\phi_0}$ whose mean energy is close to $E$, obtained in the same way, and apply a random Pauli matrix $\sigma^x$, $\sigma^y$ or $\sigma^z$ on a random site in each proposed move. This strategy gives a complete basis set, as
\begin{eqnarray}
    \frac{1}{2^N}\sum_{\substack{\mu_i=0 \\ 1\le i\le N}}^{3} \sigma^{\mu_N}_N\cdots\sigma^{\mu_1}_1 \ket{\phi_0}\bra{\phi_0} \sigma^{\mu_1}_1\cdots\sigma^{\mu_N}_N = \Id
    \label{eq:paulis}
\end{eqnarray}
for any state $\ket{\phi_0}$. The change in mean energy is $\varO(1)$ in each move, and thus this choice of basis ensures enough states for sampling.

% ====================================================================================
%          CLASSICAL SIMULATION
% ====================================================================================

\section{Classical simulation}
\label{sec:classical}

The methods that we study in this paper replace the quantum simulation of the dynamics in the quantum algorithms of~\cite{Lu2021} by classical simulations using tensor networks. The longest evolution time required in~\eqref{eq:filter_cos_sum}, which is the deciding factor for the efficiency, is determined by the width of the filter  $\delta$ as $t_{\max} = t_R \approx 2x / \delta$.
A quantum simulator should be able to deal efficiently with times $t=\varO(\mathrm{poly}(N))$, which gives access to $\delta = \Omega(\mathrm{poly}(1/N))$~\footnote{The asymptotic notations $\Omega$, $\omega$, $\varO$ and $o$ are used in this paper, which represent to be bounded from below, to dominate, to be bounded from above and to be dominated}. This should be sufficient for both~\eqref{eq:micro_filter_state} (in the case ETH is satisfied~\cite{Dymarsky2019}) and~\eqref{eq:micro_filter_tr} to converge to the microcanonical values. On a classical computer, TN techniques provide the possibility to simulate the time evolution of a local Hamiltonian~\cite{Verstraete2008,Schollwock2011,Paeckel2019tevol}, but the bond dimension required to do so can increase exponentially with time. Thus, starting from a product state, we can simulate times $t_{\max} \propto \log N$ with a bond dimension polynomial in system size, which would allow us to efficiently perform classical simulations of the algorithm for $\delta = \Omega(1 / \log N)$. For the actual implementation of the classical simulation of the dynamics, there exist several options, some of which we discuss in this section.

\subsection{Tensor network implementation}
\label{subsec:TNSimplement}
There are multiple different approaches to simulate time evolution with TN techniques (see~\cite{Paeckel2019tevol} for a recent review). Some of the most commonly used methods are based on a Suzuki-Trotter approximation of the time evolution operator. One possibility is then to repeatedly apply the approximated short time evolution steps onto a matrix product state (MPS), to obtain a representation of the time-evolved state. Alternatively, the time-evolution operator itself, $e^{-i\hat{H}t}$, can be approximated by a matrix product operator (MPO), constructed also from the iteration of trotterized steps. Therefore, we can use various techniques for the classical simulation of the quantum methods above.

Finding the MPO representation of each term $e^{-i\hat{H}t_m}$ as an MPO allows us to estimate $\mathrm{tr}(\Pd(E))$ as linear combination of the corresponding traces, which for MPOs can be computed very efficiently. This strategy will however fail as we approach the edge of the spectrum for large system sizes while considering small values of $\delta$. The reason is the extremely imbalanced distribution of the DOS $D(E)$, which becomes exponentially small when $E$ is far from the center of the spectrum. For a traceless, local and bounded Hamiltonian, in the thermodynamic limit $D(E)$ converges weakly to a Gaussian distribution with mean energy $E=0$ and width proportional to $\sqrt{N}$~\cite{Hartmann2005,Keating2015}:
\begin{eqnarray}
    \int_{-\infty}^{E_0} D(E) \dInt E \xrightarrow{N\to\infty} \int_{-\infty}^{E_0}  \frac{d^N e^{ -E^2 / 2 N \sigma_0^2}}{\sqrt{2\pi N} \sigma_0}  \dInt E,
    \label{eq:DOS_analytical}
\end{eqnarray}
where $d$ is the local Hilbert space dimension and $\sigma_0$ is some constant independent of the system size. Thus, for energies $\approx\sqrt{N}$, we expect $D(E)$ to become exponentially smaller than its value at the center. Therefore given all $\tr \left( e^{-i\hat{H}t_m} \right)$ that are reasonably precise, the ratio in~\eqref{eq:micro_filter_tr} could still not be properly achieved all through the whole spectrum: the applicable energy range will be proportional to the square root of the system size, and hence $\varO(1 / \sqrt{N})$ in energy density, a restriction we also observed in a previous work~\cite{Yang2020}.

Fortunately, this difficulty can be overcome with the importance sampling method described in \ref{subsec:q_alg}. To begin with, in this method we only need to evaluate the ratio~\eqref{eq:O_filter_state} for states for which the probability factor $\bra{\phi} \Pd (E)\ket{\phi}$ is above some threshold. In particular, when $E$ is away from the center of the spectrum, contributions from the exponentially large maximum of the DOS will be suppressed. Of course, one needs that the chosen basis $\{\ket{\phi}\}$ has enough states around the target energy $E$. Since for our numerical simulations we are free to choose any basis from which we can sample efficiently and whose states can be written as MPS, we can exploit this freedom to try to ensure this condition. A product basis minimizes the cost of the contractions and is often a good choice, since, as mentioned above, it covers an extensive window of the energy spectrum. If this is not the case, we can use any of the methods mentioned at the end of section~\ref{sec:filters} to find an MPS with small bond dimension close to the desired energy, and use it to construct a complete basis of the form~\eqref{eq:paulis}. For the cases we consider in this paper, the computational basis is already an adequate choice, sufficient to produce accurate numerical results over the full spectrum, as we illustrate in the next section.

In the Monte Carlo simulation, the time evolution can be done either at the level of the states, i.e. directly evolving the sampled state as an MPS, or at the level of the operators, i.e. approximating the evolution unitaries for each state as MPOs, storing them in memory, and using them later to do contractions with states randomly sampled from the basis. The second option has the advantage of simulating the dynamics a single time, as the same operators can be reused when doing the sampling over different states. Hence it is faster, but it also consumes much more memory to store the required MPOs~\footnote{The memory requirement can be reduced by storing only a constant fraction of the MPOs for all time steps, and combining several of them to compute the required matrix elements.}. Also, the bond dimension needed to approximate an evolution operator as an MPO is significantly larger than the one used to approximate a time-evolved MPS for the same time. As a concrete example, for system size $N$ = 80, we find bond dimension $D_{\mathrm{MPS}}=40$ to be enough for evolving MPS, and $D_{\mathrm{MPO}} \sim 100$ for storing the MPOs, for $(\delta,\alpha) \propto (1, \sqrt{N})$ and $5\times 10^4$ samples. The typical time scales taken are 1 week and 1 day, respectively with Intel\textregistered\  Xeon\textregistered\  Gold 6138 processor.

\subsection{Filter parameters and the microcanonical limit}
\label{subsec:para}
The cosine filter depends on two parameters: the width $\delta$ and the period of the filter $\alpha$. As mentioned, $\delta$ determines the maximum time we need to evolve, $t_{\max} =2 x/\delta$, while for a fixed $\delta$, $\alpha$ determines the number of terms in the expansion $R=x \alpha/\delta$.
Additionally, in the Monte Carlo algorithm we introduce a cutoff parameter $\epsilon$ and discard states for which the probability is found to be below this threshold.

In this section we discuss the significance of these parameters, as well as which choices ensure approaching the microcanonical limit. Note that the conclusions are valid for both the classical and the quantum version of the algorithms.

\subsubsection{Filter width $\delta$}
\label{subsec:para_width}

A width $\delta$ and mean energy $E_{0}$ determine the properties of the filter ensemble $\Pd(E_0)$. But to estimate the corresponding energy distribution we need to take into account the density of states. Assuming a Gaussian DOS as in~\eqref{eq:DOS_analytical}, the energy distribution of the filter ensemble will also be a Gaussian given by
\begin{eqnarray}
    \begin{aligned}
             & D_{\rhoFe{E_0}} (E) =
        D(E) \exp\left[ - \frac{(E - E_0)^2}{2\delta^2}\right]                                                                                       \\
        \sim & \exp\left(- \frac{E_0^2}{2 \gamma N\sigma_0^2} \right) \exp\left[ - \frac{\gamma}{2 \delta^2}\left(E - E_0 / \gamma\right)^2 \right],
    \end{aligned}
    \label{eq:eff_dos}
\end{eqnarray}
where $\gamma = 1 + \delta^2/N\sigma_0^2$. We omitted an energy-independent factor in~\eqref{eq:eff_dos}. It can be concluded that in the thermodynamic limit, the mean energy and width of the filtered ensemble are given by
\begin{eqnarray}
    E_{\rhoFe{E_0}} = E_0 / \gamma, \quad \Delta_{\rhoFe{E_0}} = \delta / \sqrt{\gamma}.
    \label{eq:shift_energy}
\end{eqnarray}
If we choose $\delta \propto \sqrt{N}$, the mean energy of the ensemble is shifted with respect to the parameters of the filter, as explicitly shown in Fig.~\ref{fig:trace_deltaSqrtN}. A filter width that scales as $\delta = o(\sqrt{N})$ is enough to  ensure that $E_{\rhoFe{E_0}} \to E$ and $\Delta_{\rhoFe{E_0}} \to \delta$ as $N\to\infty$. This observation is especially relevant if we are interested in approaching the microcanonical limit: in general, assuming ETH, in the thermodynamic limit a microcanonical energy shell centered at $E$ will yield the thermal values for intensive quantities at energy density $E/N$ if the width $\Delta$ satisfies $\Delta/N\to 0$. This condition is already satisfied for the filter ensemble with $\delta \propto \sqrt{N}$, which means that the expectation values will converge to the thermal ones, only at shifted energies, according to the previous argument (see Fig.~\ref{fig:trace_deltaSqrtN}).

We can similarly estimate the energy distribution of the pure state resulting from the application of the filter onto an individual state $\ket{\psi}$. There is actually a similar argument for $\Pd (E) \ket{\psi}$ if $\ket{\psi}$ is a product state, as such states also have essentially Gaussian LDOS whose widths are proportional to $\sqrt{N}$~\cite{Hartmann2004}. With a spectral decomposition $\ket{\psi} = \sum_k c_k \ket{E_k}$, where $\ket{E_k}$ are energy eigenstates, the filtered state results as
\begin{align}
    \ket{\Pd \psi}:=\sqrt{\Gamma} \Pd \ket{\psi}=\sqrt{\Gamma} \sum_k c_k e^{-\frac{(E-E_k)^2}{2 \delta^2}} \ket{E_k},
    \label{eq:filtered_state}
\end{align}
where $\Gamma = 1 / \braket{\psi | \Pd (E)^2 | \psi}$ is the normalization factor. By choosing the center of the filter at the mean energy of the state $E = \braket{\psi | \hat{H} |\psi} =: E_{\psi}$, the average energy of the filtered state does not change. Assuming the LDOS of $\ket{\psi}$ has a Gaussian form with width $\sigma_{\psi} \sqrt{N}$, where $\sigma_{\psi}$ is independent of system size, the energy variance of $\ket{\psi}$ can be estimated through substituting the sum over eigenstates by an integral over energy values with Gaussian weights. We obtain
\begin{align}
    \Delta_{\Pd\psi}^2 & :=\bra{\Pd \psi} H^2 \ket{\Pd \psi}-\bra{\Pd \psi} H \ket{\Pd \psi}^2 \nn \\
                       & \approx \frac{\delta^2}{2 + \delta^2 / N \sigma_{\psi}^2}.
    \label{eq:widthSt}
\end{align}

Again, we may want to consider how this affects approaching the microcanonical limit as the width of the filter is decreased. A major difference in this respect between the filter ensemble and the filtered state is that the second contains coherent contributions from different energy eigenstates. Thus~\eqref{eq:micro_filter_state} includes contributions from off-diagonal matrix elements in the energy basis, which only become negligible when the width of the energy distribution decreases sufficiently fast with $N$. More concretely, from canonical typicality arguments we can expect that, for non-integrable systems, the expectation value of a local observable converges to the thermal value when the energy deviation of the state decreases as a polynomial of $1/N$~\cite{Dymarsky2019}. In~\cite{Banuls2019} we observed a trend to convergence already with a slower decrease $\sim 1/\log(N)$. For these scalings of the filter width, according to~\eqref{eq:widthSt}, the width of the filtered state will scale in the same way.

\subsubsection{Period of cosine filter $\alpha$}
\label{subsec:para_alpha}

Different to the Gaussian one, the cosine filter~\eqref{eq:filter_cos} is periodic, but it remains a good approximation of the former when the argument is bounded within one period. More concretely, operators $\Pd(E)$ and $\Fd(E)$ are close to each other when $\left\lVert  \hat{H} - E \right\rVert  \le \alpha \pi / 2$. At the same time, because the number of terms that need to be evaluated in the sum~\eqref{eq:filter_cos_sum} is proportional to $\alpha$, it is convenient to choose the smallest possible value that ensures the previous property. For a local Hamiltonian, a value $\alpha \propto N$ is enough for the condition to hold for all values of $E$ within the energy spectrum. If the operator acts only on a limited energy window, a smaller value of $\alpha$ can be chosen, as long as all relevant states are almost supported in $[E - \alpha\pi/2, E + \alpha\pi / 2]$. This can be used, for instance, when the filter acts on a product state, whose energy distribution is approximately Gaussian, with support on an energy interval $\propto\sqrt{N}$. If additionally, the filter is centered near the mean energy of the state and $\delta = o(\sqrt{N})$, it is enough to choose $\alpha\propto\sqrt{N}$~\cite{Lu2021}. In practice, we find $\alpha = 3 \max(\sigma_{\phi} \sqrt{N}, \delta)$ to work well for all system sizes.

\subsubsection{Monte Carlo cutoff threshold $\epsilon$}
\label{subsec:cutoff}
For the discussions in \ref{subsec:para_alpha}, it should be ensured that the samples in Monte Carlo simulations not stepping into other energy periods of the cosine filter when $\alpha \propto \sqrt{N}$. In other words, the weights of the states whose mean energy are close the edges of $[E - \alpha\pi/2, E+\alpha\pi / 2]$ should be small enough.
A cutoff threshold $\epsilon$ can be applied to the weights of samples to improve numerical stability in Monte Carlo simulations. To be more concrete, a proposed state $\ket{\phi}$ will be directly discarded (its probability assimilated to 0) if $D_{\delta,\phi} < \epsilon$. Besides making the numerics more stable, the presence of the cutoff prevents the peak of the DOS from shifting the ensemble when we target energies near the edges,	because it restricts the visited energy range in more general cases, as we show next.

If we consider a product state basis, an individual state $\ket{\phi}$ will have mean energy $E_{\phi}$ and width $\sigma_{\phi}\sqrt{N}$. Again, assuming a Gaussian distribution, we can estimate its weight in the sum as
\begin{align}
    D_{\delta, \phi}(E) & = \int \dInt \mu \frac{1}{\sqrt{2\pi N} \sigma_{\phi}} \exp\left[ -\frac{(\mu - E_{\phi})^2}{2\sigma_{\phi}^2N} -\frac{(\mu - E)^2}{2\delta^2} \right]
    \nn                                                                                                                                                                          \\
                        & = \sqrt{\frac{\delta^2}{\delta^2 + N\sigma_{\phi}^2}} \exp\left[ - (E - E_{\phi})^2 /  ( \delta^2 + N \sigma_{\phi}^2)\right]
    \label{eq:D_delta_phi}
\end{align}
so that $D_{\delta,\phi} < \epsilon$ holds if
\begin{align}
    \left|	E-E_{\phi} \right| > \nu_{\phi} := \left[(\delta^2+N\sigma_{\phi}^2) \ln \left( \frac{\delta}{\epsilon \sqrt{\delta^2+N\sigma_{\phi}^2}}\right) \right ]^{1/2}.
\end{align}
When $\delta = \Omega(\sqrt{N})$, it follows that $\nu_{\phi} \sim \delta$ for all states in the basis. While if $\delta = \varO(\sqrt{N})$, $\nu_{\phi} = \varO(\sqrt{N})$ holds as well.  Hence for any $\delta = o(N)$, the cutoff $\epsilon$ itself can restrict the sampling space within an energy interval of width $o(N)$ that screens the peak of the DOS.

\begin{table*}
    \centering
    \begin{tabular}{|c|c|c|c|c|c|c|c|}
        \hline
              & Method                                            & $\delta$               & $\alpha$           & $t_{\max}$             & $R$                & Applicable energy range        \\
        \hline
        (I)   & Spectrum filtering, direct trace                  & $o(\sqrt{N})$          & $N$                & $\omega(1 / \sqrt{N})$ & $\omega(\sqrt{N})$ & $\varO(\sqrt{N})$              \\
        \hline
        (IIa) & \multirow{2}{*}{Spectrum filtering, Monte Carlo } & $\varO(\sqrt{N})$      & $N$                & $\Omega(1 / \sqrt{N})$ & $\Omega(\sqrt{N})$ & \multirow{3}{*}{Full spectrum} \\
        \cline{1-1}\cline{3-6}
        (IIb) &                                                   & $\varO(1)$             & $\Omega(\sqrt{N})$ & $\Omega(1)$            & $\Omega(\sqrt{N})$ &                                \\
        \cline{1-6}
        (III) & State filtering                                   & $\mathrm{poly}(1 / N)$ & $\Omega(\sqrt{N})$ & $\mathrm{poly}(N)$     & $\mathrm{poly}(N)$ &                                \\
        \hline
    \end{tabular}
    \caption{Possible choices of the filter parameters $(\delta,\alpha)$ that ensure approaching the microcanonical values in the thermodynamic limit for the various methods, and corresponding maximum evolution time and number of steps. }
    \label{table:para}
\end{table*}

\subsubsection{Choosing the parameters for microcanonical values}
\label{subsec:params}
According to the discussions above, we can summarize in Table~\ref{table:para} some possible choices of parameters for the various algorithms, such that we obtain convergence to thermal values in the thermodynamic limit.
The table shows the scaling with system size of $\alpha$ and $\delta$, as well as the resulting cost (in terms of maximum evolution time and number of evolutions to run) and the energy range where the methods are applicable.
The fastest method [(I) in the table] is directly computing~\eqref{eq:micro_filter_tr} by taking the traces of the evolution operators approximated as MPOs, but, as discussed in section~\ref{subsec:TNSimplement}, it is only applicable in an energy interval of width proportional to $\sqrt{N}$ around the center of the spectrum. With Monte Carlo sampling (II), in contrast, it is possible to reach the whole spectrum. The filter width required to obtain convergence to thermal values in the thermodynamic limit should scale at most as $\varO(\sqrt{N})$. A larger width corresponds to a shorter evolution time $t_{\max}$, and hence a smaller bond dimension required for the MPS or MPO, but it also shows slower convergence to the thermal values as the system size is increased. We thus show two possible choices (IIa) and (IIb), both of which we explore numerically, using different approaches for time evolution, in \ref{subsec:res_filter_ens}. In (IIa), a cutoff threshold is applied in the Monte Carlo simulations to avoid the energy shift due to DOS when $\delta \propto \sqrt{N}$. Finally, when filtering a state (III), the time $t_{\max}$ required to approach microcanonical values is polynomial in the system size, which means that with TN techniques we will be able to extract microcanonical values with this method only for small system sizes.
For this last method, however, the achievable width can be optimized by applying the algorithm on a state with reduced energy width. We present a way to implement this improvement by using MPS obtained after a variational minimization of the variance.

% ====================================================================================
%          RESULTS
% ====================================================================================

\section{Results}
\label{sec:results}

To demonstrate and  benchmark the various methods described above, we apply them
to a quantum Ising chain with open boundary conditions,
\begin{eqnarray}
    \hat{H}_{\mathrm{Ising}} = J \sum_{i=1}^{N-1} \sigma_i^z \sigma_{i+1}^z + \sum_{i=1}^{N} (g\sigma_i^x + h\sigma_i^z).
    \label{eq:Ising}
\end{eqnarray}
The model is integrable if either $g=0$ or $h=0$.
Here we choose a particular set of parameters $(J,g,h)=(1,-1.05,0.5)$ far from integrability~\cite{Kim2013ballistic}.
In the thermodynamic limit, the corresponding energy density lies in the interval $E/N\in[-1.33,1.72]$. For the observable, we focus on the average magnetization
\beq
\hat{m}_z = \sum_{i=1}^{N} \sigma_i^z / N.
\label{eq:mz}
\eeq

\subsection{Filter ensemble}
\label{subsec:res_filter_ens}

We start by illustrating the performance of the Monte Carlo algorithm to estimate expectation values in the filter ensemble~\eqref{eq:rhoF} at all values of energy. Since the largest time we need to simulate is $t_{\max}\propto 1/\delta$, the classical simulation can efficiently treat widths $\delta= \varO(1/\log N)$ and larger.

For the numerical benchmarking, we choose widths $\delta\propto\sqrt{N}$ and $\delta=\mathrm{const}$, which, according to the discussions above, are enough to approach the microcanonical values in the thermodynamical limit [see (II) in table~\ref{table:para}]. In the non-integrable model we consider, the values are thus expected to converge to the thermal ones. Thus, we can compare the results of the algorithm with the exact values in thermal equilibrium at the corresponding energies, which we can compute independently using standard TN techniques~\cite{Stoudenmire2010}.

The calculations can be done using different options for the TN evolution (section~\ref{subsec:TNSimplement}). As long as the results are converged in bond dimension, both approximating the evolution operators as MPO or the individual evolved states as MPS are valid strategies, and we show results obtained with both of them.

\subsubsection{MPO version of Monte Carlo simulation}
\label{subsec:res_MC_mpo}

\begin{figure}[t]
    \centering
    \includegraphics[width=0.48\textwidth]{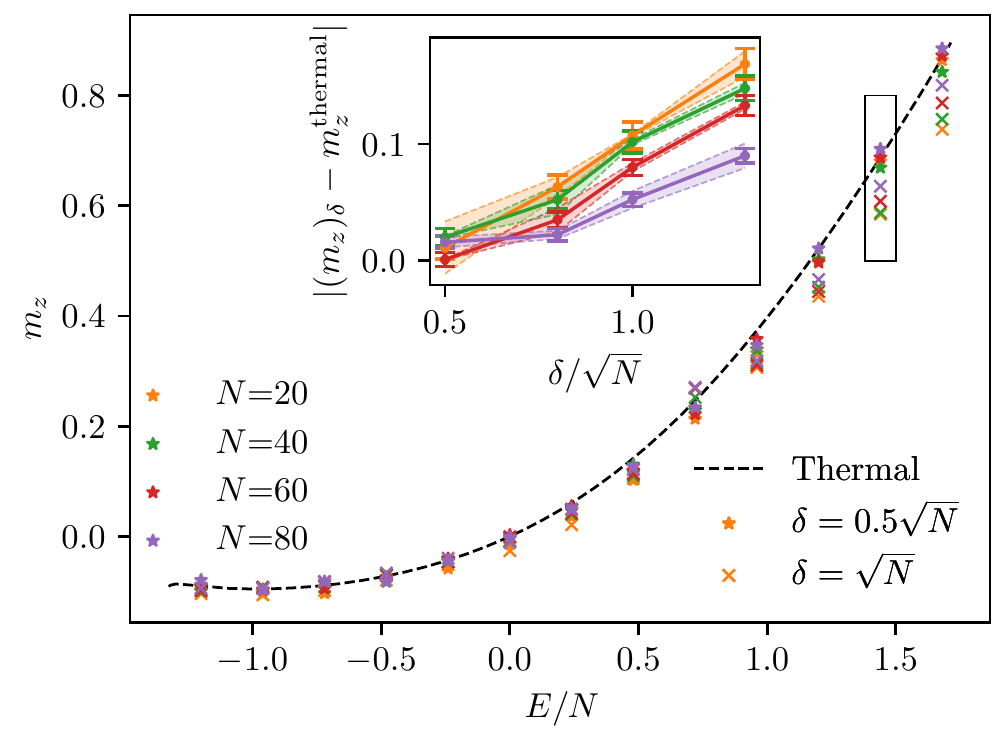}
    \includegraphics[width=0.23\textwidth]{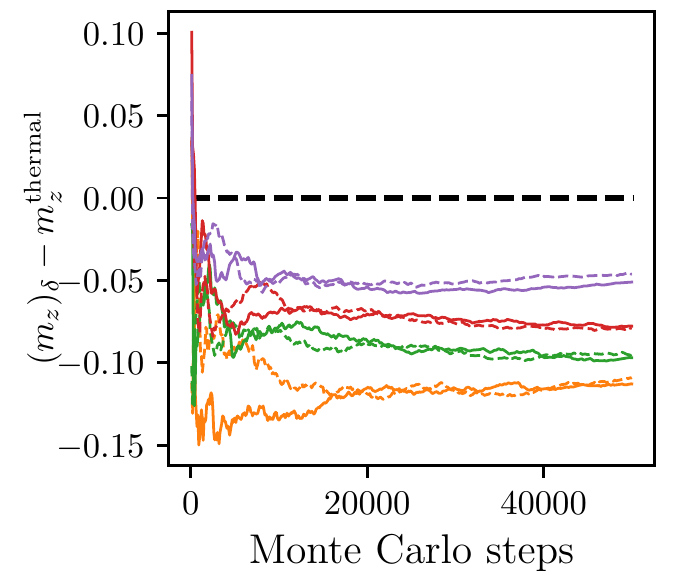}
    \includegraphics[width=0.23\textwidth]{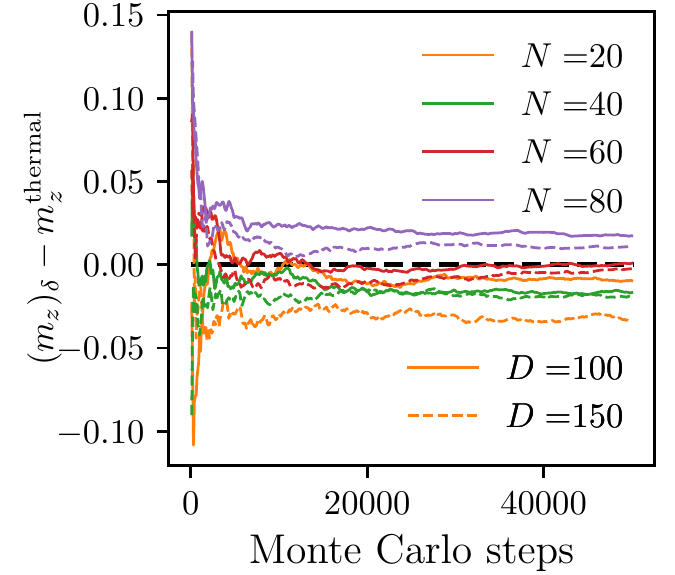}
    \caption{
        Magnetization~\eqref{eq:mz} in the non-integrable Ising chain~\eqref{eq:Ising} with $(J,g,h)=(1,-1.05,0.5)$, computed by the MPO version of the Monte Carlo method with parameters $(\delta \propto \sqrt{N},\,\alpha = \sqrt{J^2+g^2+h^2} N)$ and a cutoff (see main text). The sample size is $5\times 10^4$ for each energy and $\delta$. \textbf{Upper panel:} Convergence over the whole spectral range.
        The black dashed line is the thermal value for $N = 80$. Data points are $(m_z)_{\delta}$, where different colors stand for different system sizes and different shapes for different $\delta$. \textbf{The inset} plots the difference with respect to the thermal value as a function of $\delta/\sqrt{N}$ at the point $E/N=1.44$ (indicated by a box in the main plot), with error bars indicating the variance of the Monte Carlo sampling, while the shadowed region represents errors from finite MPO bond dimensions ($D = 100$ vs. $D = 150$).
        \textbf{Lower panels:} Convergence with the sample size at $E / N = 1.44$ for $\delta = \sqrt{N}$ (left) and $\delta= 0.5\sqrt{N}$ (right).
    }
    \label{fig:MC_MPO_sqrtN_cut}
\end{figure}

Figure~\ref{fig:MC_MPO_sqrtN_cut} demonstrates the success of the method to find expectation values in the filter ensemble, for system sizes up to $N=80$. In particular, for this plot, we chose to simulate and store the MPOs for all evolution operators before realizing the sampling over the computational basis. For the filter, we used filter parameters $(\delta, \alpha) \propto (\sqrt{N}, N)$, which, as argued in section~\ref{subsec:para}, in a generic case is enough for the observable to converge, in the thermodynamic limit, to the thermal expectation value if introducing a cutoff threshold (section~\ref{subsec:cutoff}). We choose to sample over the computational basis and the cutoff threshold $\epsilon = 10^{-4} D_{\delta,\phi_0}(E)$, where $\ket{\phi_0}$ is the initial state in the Monte Carlo simulation, obtained by minimizing $\braket{\phi| (\hat{H} - E)^2 | \phi}$ for $\ket{\phi}$ in the basis set.

As shown in the upper panel of fig.~\ref{fig:MC_MPO_sqrtN_cut}, the results clearly converge to the thermal value as the system size $N$ is increased or $\delta$ is reduced. The inset shows explicitly this convergence for energy density $E/N=1.44$, relatively close to the edge of the spectrum. Errors have two main sources, which are shown in this plot: the statistical error from the Monte Carlo sampling (error bars) and the truncation error from the finite bond dimension of the MPO (shown as shadowed region). Only for the largest system size $N=80$ and smallest width $\delta = 0.5\sqrt{N}$ we observe a small discrepancy, but compatible with our estimated errors from both sources.

The lower panels of fig.~\ref{fig:MC_MPO_sqrtN_cut} show explicitly the convergence of the Monte Carlo sampling at the same energy density $E/N=1.44$, for various system sizes and bond dimensions, and for two different values of the width. In all these cases we observe that, after 50000 steps, the results are practically converged, even though some fluctuations can be appreciated.

\subsubsection{MPS version of Monte Carlo simulation}
\label{subsec:res_MC_mps}

To illustrate the performance of the algorithm when individual states, rather than operators, are evolved, we choose a narrower filter width, which should lead to values closer to the thermal ones, while still being reachable by classical simulations. Notice, however, that similar values could have been obtained with the MPO option, at a different cost in memory and time.

In particular, considering a constant value $\delta=\varO(1)$, independent of system size, requires evolution until a constant time, which (for not too small values of $\delta$) can be efficiently simulated using TN. Thus we choose parameters $(\delta,\alpha)\propto(1, \sqrt{N})$ [(IIb) in Table~\ref{table:para}], and explore two values of the energy density, one near the center of the spectrum ($E/N=0.72$) and one close to the edge ($E/N=1.44$), and both within the reach of product states from the computational basis, which we take again as our sampling basis.

\begin{figure}[t]
    \centering
    \includegraphics[width=0.23\textwidth]{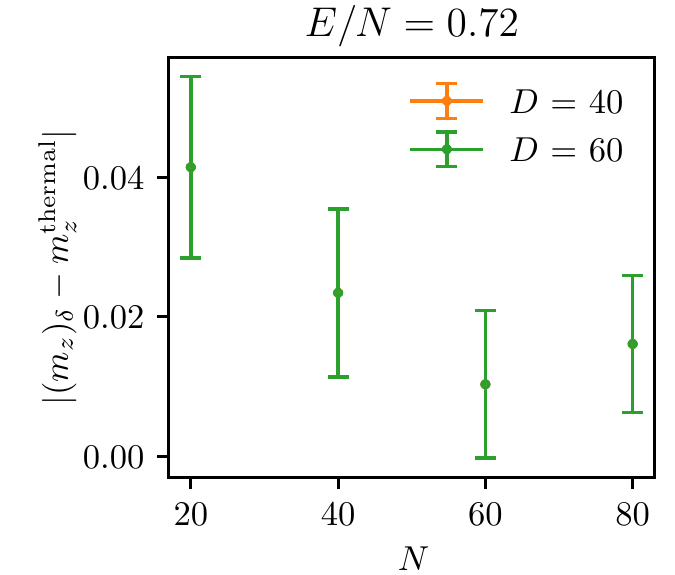}
    \includegraphics[width=0.23\textwidth]{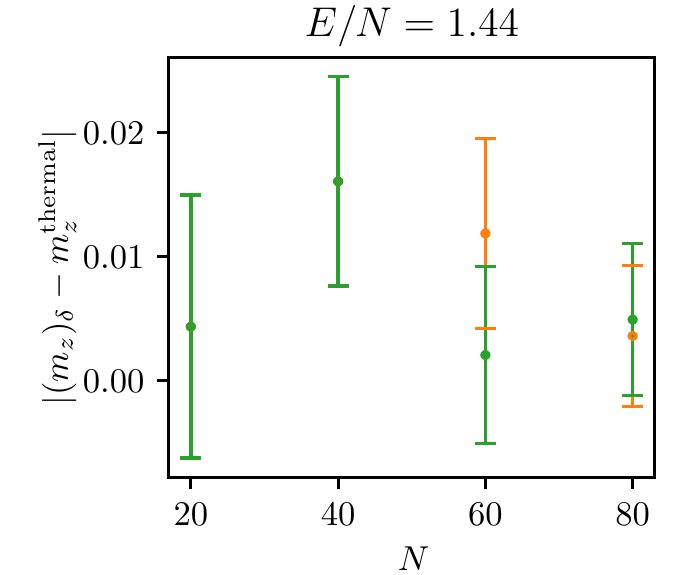}
    \includegraphics[width=0.23\textwidth]{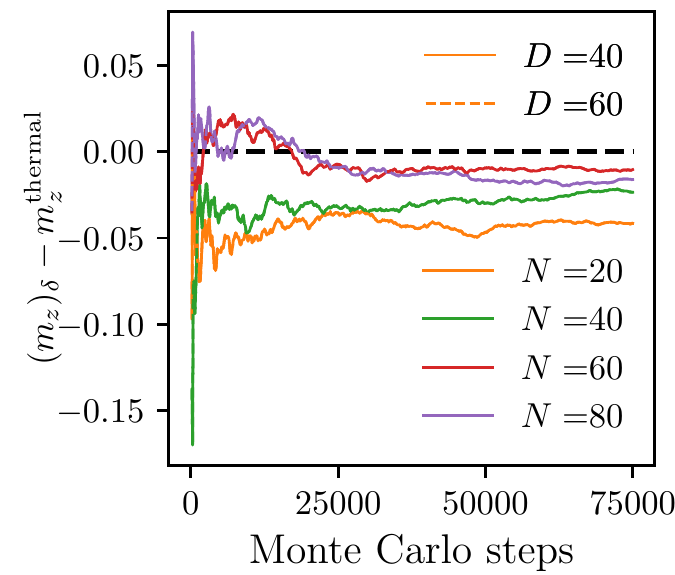}
    \includegraphics[width=0.23\textwidth]{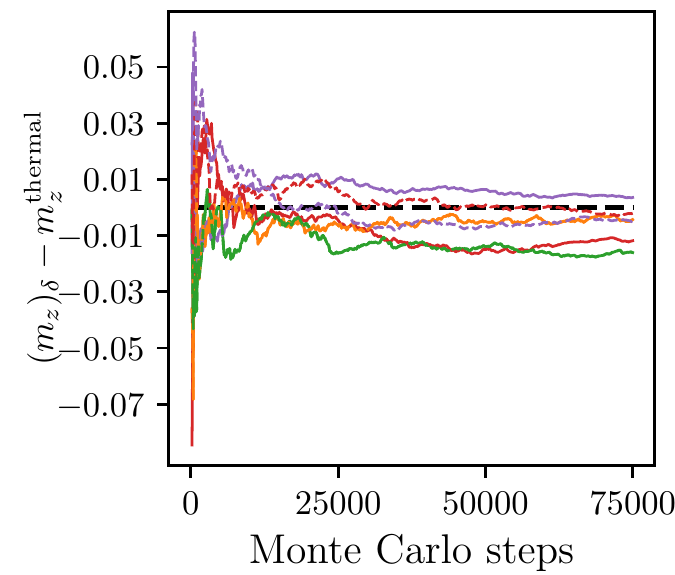}

    \caption{
        Magnetization~\eqref{eq:mz} obtained by the MPS version of the Monte Carlo method for $(\delta = 1,\, \alpha = 6\sqrt{N})$ and no cutoff, for two values of the energy density, $E/N=0.42$ (left) and $E/N=1.44$ (right). \textbf{Upper panels:} Difference between $(m_z)_{\delta}(E)$ and thermal value as a function of the system size for bond dimensions $D=40,\ 60$. The error bars correspond to Monte Carlo fluctuations. \textbf{Lower panels:} convergence of Monte Carlo sampling with the number of steps. The results have converged for bond dimension $D=40$ in the left plots, and hence the solid and dashed lines are on top of each other.}
    \label{fig:MC_MPS_1_edge}
\end{figure}

Results are shown in Fig.~\ref{fig:MC_MPS_1_edge}. In the upper panels, we plot the difference between $(m_z)_{\delta}$ and the corresponding thermal value as a function of the system size, for two values of the bond dimension,
with error bars indicating the statistical error.
We find that, within error bars, the distance to the thermal value decreases as the system size grows, with a relative difference smaller than $0.5\%$ for $N=80$ and $E/N=1.44$.

In the lower panels of Fig.~\ref{fig:MC_MPS_1_edge}, we again show explicitly the convergence of the Monte Carlo sampling. Note that here the same seed for randomization was used for different bond dimensions, so the fact that the solid and dashed lines (representing $D=40$ and $D=60$ respectively) are on top of each other indicates the convergence with regard to bond dimension already at $D=40$.

As illustrated above, the combination of short-time dynamics simulation and sampling provides a powerful method to compute the expectation values in the filter ensemble, as long as the basis contains vectors with substantial weight in the energy region of interest. For the computational basis that we have used in the examples, mean energies lie in the interval $E/N\in[-1, 1.5]$. According to~\eqref{eq:D_delta_phi}, the weights of basis states $D_{\delta,\phi}$ will decay exponentially with the system size for a fixed energy density $E / N$ outside this interval. Fig.~\ref{fig:MC_MPO_sqrtN_cut} shows that, for system size $N=80$, the Monte Carlo sampling with computational basis remains valid at $E/N=1.68$ (rightmost point) and $E/N=-1.2$ (leftmost point), much closer to the edges of the spectrum, so that we do not need to resort to the Pauli basis mentioned in \ref{subsec:q_alg}. Using this basis may however become necessary as we keep increasing the system size, or if we consider other models or higher dimensions.
\%it to finally fail as keeping increasing the system size, or when we deal with other models with larger gap between product states and the ground (or maximally excited) state. In these cases the Pauli basis mentioned in \ref{subsec:q_alg} will be applicable.

\subsubsection{Exploring the center of the spectrum without sampling}
\label{subsec:res_traces}

\begin{figure}[t]
    \includegraphics[width=.48\textwidth]{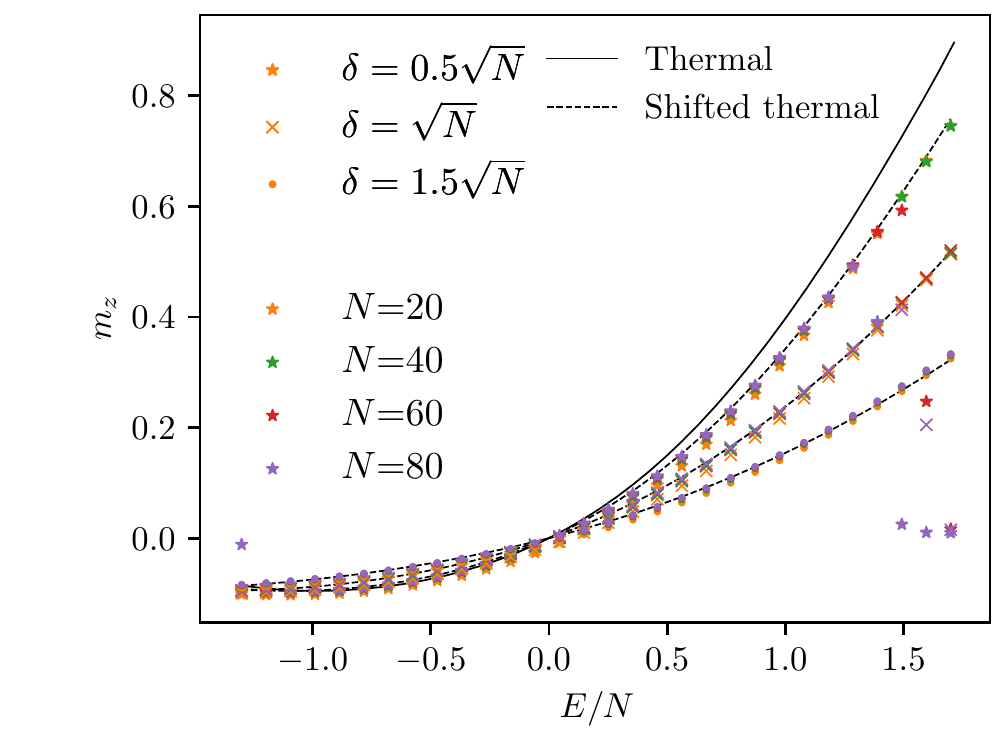}
    \includegraphics[width=.48\textwidth]{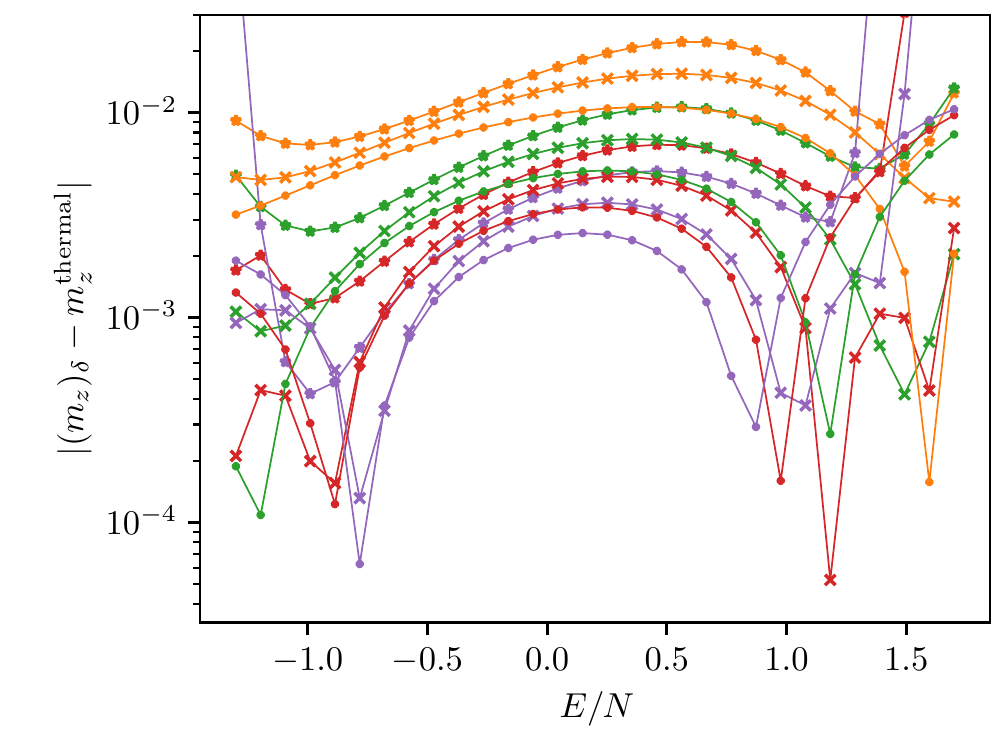}
    \caption{Magnetization~\eqref{eq:mz} obtained by MPO simulation of the evolution operators and direct evaluation of the ratio~\eqref{eq:micro_filter_tr}. \textbf{Upper panel:} computed magnetization as a function of the energy density $E/N$ at the center of the filter for system sizes up to $N=80$
        and several widths.
        Over most of the spectral range the results converge to thermal values at the shifted energy depending on $\delta$~\eqref{eq:shift_energy} (dashed lines) instead of the thermal value at $E/N$.
        At the edges of the spectrum this fails, due to the exponentially decaying DOS in these regions, as discussed in \ref{subsec:para_width}.
        \textbf{Lower panel:} difference between the computed magnetization and the shifted thermal values.
        All results were obtained using bond dimension $D=200$ for the MPOs.
    }
    \label{fig:trace_deltaSqrtN}
\end{figure}

The fastest alternative to evaluate~\eqref{eq:micro_filter_tr} with TN simulations is to directly evaluate numerator and denominator from traces of the evolution MPOs, without the sampling iteration [(I) in Table~\ref{table:para}].	As discussed above, this is only feasible in the central region of the spectrum, over a width $\propto \sqrt{N}$, before the density of states becomes exponentially small. If the filter width is not much smaller than this scale, the mean energy of the filter ensemble will be effectively shifted towards the maximum of the DOS, as explicitly computed in sec.~\ref{subsec:para_width}.

Figure~\ref{fig:trace_deltaSqrtN} illustrates the behavior of this alternative for $(\delta, \alpha) \propto (\sqrt{N}, N)$. The upper plot shows the results obtained for the magnetization $m_z$ for various system sizes and filter widths, as a function of the energy density corresponding to the center of the filter. Because of the shift discussed above, the results do not converge to the thermal ones (indicated by the solid line) at that energy, but at a shifted value according to~\eqref{eq:shift_energy} (indicated by a dashed line for each $\delta$).

We observe that, while in the central part of the spectrum better convergence is observed as $\delta$ decreases or $N$ increases, near the edge of the spectrum the method fails to give the correct microcanonical values, especially for larger systems and smaller widths, as then it becomes sensitive to the exponentially smaller density of states. This is also visible in the lower panel of the figure, where we plot the difference between the computed values $O_{\delta}(E)$ ad the thermal ones at the corresponding shifted energies. According to~\eqref{eq:eff_dos}, the denominator of~\eqref{eq:micro_filter_tr}, $\tr \left[\Pd (E)\right]$ should scale as $\exp(-E^2 / 2\gamma N\sigma_0^2)$, indicating the range of energy densities for which the method actually converges shrinks as $1 / \sqrt{N}$.

\subsection{Filtered pure state}
\label{sec:state_filtering}
As discussed in section~\ref{subsec:para}, we can apply the filter on a state to decrease its energy variance and, in the generic case, obtain convergence to the microcanonical properties. Also in this case, the largest time that needs to be simulated is $t_{\max} = 2x / \delta$, which can be done efficiently by TN simulations if the width is at least $\varO(1/\log N)$. But, in contrast to the calculations for the filter ensemble, diagonal in the energy basis, a much smaller $\delta$ is required in this case to approach microcanonical values in the thermodynamic limit. More concretely, $\delta = \varO(\mathrm{poly}(1/N))$ should be enough, but this requires $t_{\max} = \Omega(\mathrm{poly}(N))$, and thus a bond dimension that increases exponentially with $N$. Additionally, when filtering a product state, the total number of terms to probe will be $(2R+1)^2$, where $R = x\alpha/\delta$, which grows at least as $N^3$.

\begin{figure}[ht]
    \centering
    \includegraphics[width=0.23\textwidth]{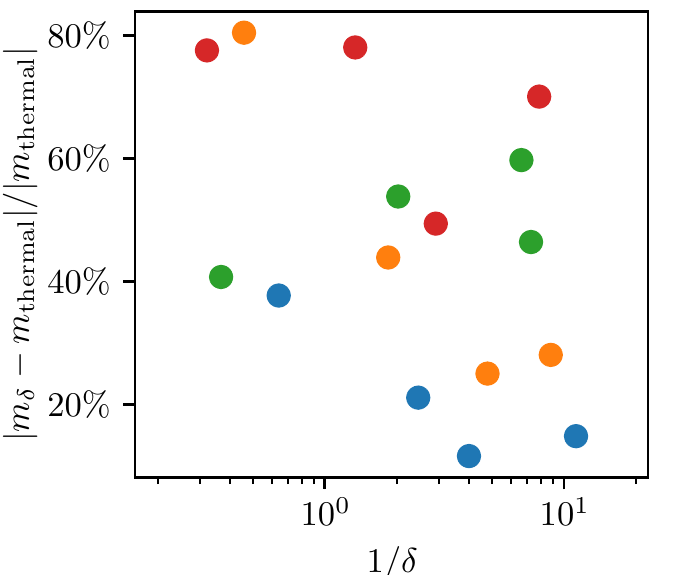}
    \includegraphics[width=0.23\textwidth]{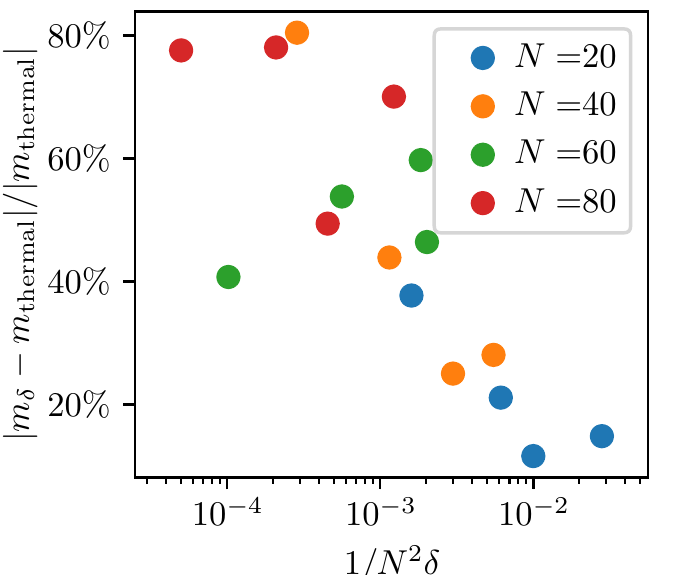}
    \caption{Results computed by filtering states. $E/N=0.72$. The filtered states are MPS of bond dimension 1, 2, 5 and 10. We compare the results of different system sizes with $\delta  \propto 1$ in the left figure and $\delta \propto 1 / N^2$ in the right.}
    \label{fig:MPS_edge}
\end{figure}

One way to mitigate the second problem is to apply the filter on a state with already reduced energy variance. This allows us to choose a smaller period $\alpha$ and correspondingly keep a more moderate value of $R$ and to test the strategy for moderate sizes. We implement this strategy using as initial states MPS with a given bond dimension, found by variationally minimizing $\left(\hat{H} - E\right)^2$ at the value of $E$ we are interested in. To test this strategy, we have targeted a value $E/N=0.72$ for system sizes $20\leq N\leq 80$, and obtained initial MPS with reduced widths from the minimization of $(H-E)^2$ with bond dimensions $D_0\in\{1,\,2,\,5,\,10\}$. For each one of this states, we compute the width $\sigma_D$, and then apply a filter with parameters $\delta=\sigma_D/2\sqrt{N}$ and $\alpha=3 \sigma_D$. We show the results in Fig.~\ref{fig:MPS_edge}. To analyze the convergence towards the thermal value as the width decreases, we plot the relative error of the magnetization with respect to the thermal one as a function of $1/\delta$ (left panel) and $1/(N^2 \delta)$ (right panel) for each system size. The first case shows no clear scaling laws, which indicates that $\delta = \varO(1)$ is not enough for convergence in the thermodynamic limit. The right panel, instead, exhibits a trend to convergence for $\delta = \varO (1 / N^2)$, even though numerically it becomes difficult to reach the lower right corner for large systems.

% ====================================================================================
%          DISCUSSION
% ====================================================================================

\section{Summary and Discussion}
\label{sec:discussion}

We have presented a quantum-inspired classical method, based on a TNS simulation of the quantum-assisted Monte Carlo algorithm proposed in~\cite{Lu2021}, that allows us to compute microcanonical and diagonal values for quantum many-body systems. Our method estimates broadened spectral functions, which takes the form of the trace of an energy filter operator, or a product of the latter with an observable, via sampling over time-evolved product states. Because the longest required time is proportional to the inverse filter width, that is very short and easy to simulate with tensor networks, it allows us to find expectation values in a diagonal ensemble which would only be reached after a much longer time evolution from an initial product state. While filter widths $\varO(\sqrt{N})$ are enough to find the diagonal ensemble values of generic product states, we can also reach energy filters of constant widths, as these only require $O(1)$ evolution times. These scalings are enough to obtain convergence to thermal equilibrium in the thermodynamic limit, in the generic case.

We have benchmarked the algorithm on the non-integrable Ising chain, for sizes up to $N=80$ sites (far beyond the reach of exact diagonalization), and we have checked different choices of the parameters that affect the efficiency and applicability of the method. In particular, we explicitly show diagonal expectation values for Gaussian ensembles of width $\varO(\sqrt{N})$, obtained with low computational cost over the whole range of energies, and observe their convergence towards the thermal equilibrium values. Reducing further the width, we obtain microcanonical expectation values with high precision.

We can also classically simulate the provably efficient quantum algorithm in~\cite{Lu2021}, in which the filter is applied on a fixed initial state. If we want to use this method to explore the microcanonical properties, nevertheless, the filter width needs to decrease with the system size in order to guarantee convergence in the thermodynamic limit, which results in increasing times and an exponentially growing bond dimension. We have however optimized the procedure by choosing as initial states MPS with minimal variance. In this way, we can run the algorithm and observe convergence for reasonably sized systems.

The results shown in this paper demonstrate the potential of the algorithm. Accessing the microcanonical values would be helpful to investigate all sorts of out-of-equilibrium quantum many-body behavior, for instance many-body localization~\cite{Abanin2018} and quantum scars~\cite{Turner2018a}, for large systems. Although we have only implemented it for a translationally invariant spin chain with short-range interactions, the method can be easily extended to systems with disorder, or long-range interactions, and also to bosonic or fermionic systems. We have recently learned that A. Schuckert et al. are using a related method to study long-range interacting models~\cite{Schuckert2022}.

In principle, the same method can be also extended to higher dimensional systems, using PEPS (projected entangled pairs states)~\cite{Verstraete2004b}, which is a promising possibility, given the absence of numerical methods for extracting microcanonical expectation values in that case, beyond exact diagonalization. The numerical challenge is then higher, due to the larger computational cost of the corresponding algorithms, and determining the preferable implementation option should be analyzed. Also further extensions are possible that consider other filter functions or combinations of filters.

% Acknowledgments
\begin{acknowledgments}
    This work was partly supported by the Deutsche Forschungsgemeinschaft (DFG, German Research Foundation) under Germany's Excellence Strategy -- EXC-2111 -- 3908148688, by the European Union through the ERC grant QUENOCOBA, ERC-2016-ADG (Grant No. 742102) and by the German Federal Ministry of Education and Research (BMBF) through the funded project EQUAHUMO (Grant No. 13N16066) within the funding program quantum technologies - from basic research to market, in association to the Munich Quantum Valley.
\end{acknowledgments}

\bibliographystyle{apsrev4-2}
\bibliography{filter_ref}
\end{document}